\documentstyle[mprocl]{article}

\input epsf

\begin{document}

\title{TESTS OF STRONG-FIELD GRAVITY AND GRAVITATIONAL\\
RADIATION DAMPING IN BINARY-PULSAR SYSTEMS}

\author{G. ESPOSITO-FARESE}

\address{CPT, CNRS Luminy,
Case 907, F 13288 Marseille cedex 9, France\\
and DARC,
Observatoire de Paris-Meudon, F 92195 Meudon cedex, France
\\E-mail: gef@cpt.univ-mrs.fr
}

\maketitle\abstracts{This talk reviews the constraints imposed by
binary-pulsar data on gravity theories, and notably on
``scalar-tensor'' theories which are the most natural alternatives to
general relativity. Because neutron stars have a strong gravitational
binding energy, binary-pulsar tests are qualitatively different from
solar-system experiments: They have the capability of probing models
which are indistinguishable from general relativity in weak
gravitational field conditions. Besides the two most precise
binary-pulsar experiments, in the systems B1913+16 and B1534+12, we
also present the results of the various ``null'' tests of general
relativity provided by several neutron star--white dwarf binaries,
notably those of gravitational radiation damping.}

Binary pulsars allow us to test the strong-field regime of gravity
because neutron stars are compact objects: Their mass to radius
ratio is of order $Gm/Rc^2 \sim 0.2$, as compared to $\sim
10^{-6}$ for the Sun. The best way to illustrate the various tests
available is to plot their constraints on a generic class of
scalar-tensor theories,\cite{nonpert} defined by the following
action:
\begin{equation}
S = \int d^4 x \sqrt{-g} \Bigl( R - 2 g^{\mu\nu} \partial_\mu \varphi
\partial_\nu \varphi \Bigr) + S_{\rm matter}\Bigl[{\rm matter};
e^{2\alpha_0\varphi + \beta_0 \varphi^2}g_{\mu\nu}\Bigr]\ ,
\end{equation}
where $\alpha_0$ and $\beta_0$ are two constants characterizing the
coupling of matter to the scalar field. The origin of Fig.~1
($\alpha_0 = \beta_0 = 0$) corresponds to general relativity;
the vertical axis ($\beta_0 = 0$) to Brans-Dicke theory with a
parameter $2\omega + 3 = 1/\alpha_0^2$; and the horizontal one
($\alpha_0 = 0$) to theories which are perturbatively equivalent to
general relativity, {\it i.e.}, strictly indistinguishable from it in
the weak-field conditions of the solar system.

Lunar Laser Ranging experiments\,\cite{etaLLR} and observations of
Mercury perihelion shift
\ exclude the region lying above the thin solid line of Fig.~1, while
Very Long Baseline Interferometry\,\cite{gammaVLBI} imposes the limit
$|\alpha_0| < 1.4 \times 10^{-2}$ materialized by a thin horizontal
line. This figure confirms that solar-system experiments do not
impose any constraint on $\beta_0$ if $\alpha_0$ is small enough.

On the other hand, binary-pulsar data impose\,\cite{nonpert} $\beta_0
> -4.5$ even for a vanishingly small $\alpha_0$. In terms of the
Eddington parameters $\beta^{\rm PPN}$ and $\gamma^{\rm PPN}$, which
are both consistent with 1 in the solar system, this inequality reads
$(\beta^{\rm PPN}-1) / (\gamma^{\rm PPN}-1) < 1.1$. This result comes
mainly from the Hulse-Taylor binary pulsar,\cite{1913} which
excludes the region lying above the bold line labeled ``1913+16''.

The next strongest constraint is imposed by neutron star--white dwarf
binaries like PSR 0655+64, which excludes the region outside the
two solid lines of Fig.~1. Indeed, in scalar-tensor theories, such
dissymmetrical systems emit strong dipolar gravitational waves
($\propto 1/c^3$), generally inconsistent with the small observed
value of their $\dot P_b$ (rate of change of orbital period). The
quadrupolar radiation ($\propto 1/c^5$) predicted by general
relativity is consistent with experimental data.

The binary pulsar\,\cite{1534} PSR 1534+12 excludes the region lying
to the left and top of the dashed line. This is a slightly weaker
constraint than with the previous systems, but this test is
nevertheless very important, because it does not depend on the
radiative structure of the theory.

Scalar-tensor theories also predict a violation of the strong
equivalence principle (SEP), which causes a gravitational analogue
of the Stark effect on the orbit of dissymmetrical
binaries.\cite{Stark,Wex00} The small observed eccentricities of such
systems constrain the theories to lie between the two (approximate)
dotted lines.

Similar analyses can also be performed to test local Lorentz
invariance of gravity and conservation laws, and notably the ``PPN''
parameters\,\cite{alpha1PSR,Wex00} $\alpha_1$
and\,\cite{alpha3,Wex00} $\alpha_3$.

\begin{figure}
\begin{center}\leavevmode\epsfbox{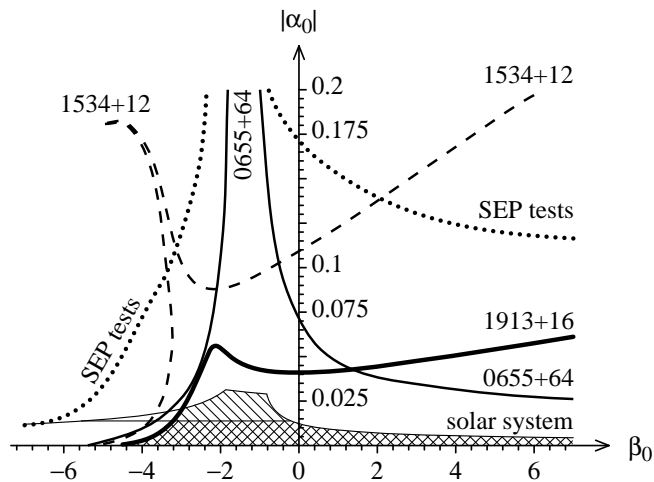}\end{center}
\caption{Solar-system \& binary-pulsar constraints on
generic scalar-tensor theories. The hatched region is allowed by
all the tests. The doubly hatched one is also consistent
with VLBI data.$^3$
\label{fig1}}
\end{figure}

\end{document}